\documentclass[preprintnumbers,amsmath,11pt,amssymb,floatfix,
superscriptaddress,nofootinbib]{revtex4}
\usepackage{graphicx}
\usepackage{epsfig}
\usepackage{bm}
\usepackage{amsfonts}

\input epsf

\begin{document}
\title{\Large Generalized Second Law of Thermodynamics for FRW Cosmology with
 Power-Law Entropy Correction}

\author{{\bf Ujjal Debnath}}
\email{ujjaldebnath@yahoo.com , ujjal@iucaa.ernet.in}
\affiliation{Department of Mathematics, Bengal Engineering and
Science University, Shibpur, Howrah-711 103, India.}

\author{\bf Surajit Chattopadhyay}
\email{surajit_2008@yahoo.co.in,
surajit.chattopadhyay@pcmt-india.net} \affiliation{Department of
Computer Application (Mathematics Section), Pailan College of
Management and Technology, Bengal Pailan Park, Kolkata-700 104,
India.}

\author{\textbf{Ibrar Hussain}}
\email{ibrar.hussain@seecs.nust.edu.pk} \affiliation{School of
Electrical Engineering and Computer Science (SEECS), National
University of Sciences and Technology (NUST), H-12, Islamabad,
Pakistan.}

\author{\textbf{Mubasher Jamil}}
\email{mjamil@camp.nust.edu.pk , jamil.camp@gmail.com}
\affiliation{Center for Advanced Mathematics and Physics (CAMP),
National University of Sciences and Technology (NUST), H-12,
Islamabad, Pakistan.}

\author{\textbf{Ratbay Myrzakulov}}
\email{rmyrzakulov@csufresno.edu, rmyrzakulov@gmail.com}
\affiliation{Eurasian International Center for Theoretical
Physics, Eurasian National University, Astana 010008, Kazakhstan.}

\date{\today}

\begin{abstract}\textbf{Abstract:}
In this work, we have considered the power law correction of entropy
on the horizon. If the flat FRW Universe is filled with the $n$
components fluid with interactions, the GSL of thermodynamics for
apparent and event horizons have been investigated for equilibrium
and non-equilibrium cases. If we consider a small perturbation
around the de Sitter space-time, the general conditions of the
validity of GSL have been found. Also if a phantom dominated
Universe has a polelike type scale factor, the validity of GSL has
also been analyzed. Further we have obtained constraints on the
power-law parameter $\alpha$ in the phantom and quintessence
dominated regimes. Finally we obtain conditions under which GSL
breaks down in a cosmological background.
\end{abstract}

\maketitle
\newpage

\section{\bf Introduction}

In Einstein gravity, the evidence of a connection between
thermodynamics and Einstein field equations was first discovered in
\cite{Jac} by deriving the Einstein equation from the
proportionality of entropy and horizon area together with the first
law of thermodynamics $\delta Q = T dS$ in the Rindler spacetime.
The horizon area of black hole is associated with its entropy, the
surface gravity is related with its temperature in black hole
thermodynamics \cite{Bek,Hawk}. The Friedmann equation in a
radiation dominated Friedmann-Robertson-Walker (FRW) Universe can be
written in an analogous form of the Cardy-Verlinde formula, an
entropy formula for a conformal field theory \cite{Ver}. As is well
known, event horizons, whether of black holes or cosmological, mimic
black bodies and possess a non-vanishing temperature and entropy,
the latter obeying the Bekenstein-Hawking entropy formula
\cite{Gibb,Dav} $S=\frac{A}{4}$ ($c=G=\hbar=1$), where $A=4\pi
R_{h}^{2}$ is the area of the horizon, $R_{h}$ is the radius of the
horizon and $G$ is the Newton's gravitational constant. The first
law of thermodynamics for the cosmological horizon is given by
$-dE=TdS$, where $T=\frac{1}{2\pi R_{h}}$ is the Hawking temperature
\cite{Cai1, Gong}.

Recently, it was demonstrated that cosmological apparent horizons
are also endowed with thermodynamical properties, formally identical
to those of event horizons \cite{Cai2}. In a spatially flat de
Sitter space–time, the event horizon and the apparent horizon of the
Universe coincide and there is only one cosmological horizon. When
the apparent horizon and the event horizon of the Universe are
different, it was found that the first law and generalized second
law (GSL) of thermodynamics hold on the apparent horizon, while they
break down if one considers the event horizon \cite{Wang}. Recently,
it has been demonstrated that if the expansion of the Universe is
dominated by phantom energy, black holes will decrease their mass
and eventually disappear altogether \cite{pavon,jamil123}. This
means a threat for the GSL as these collapsed objects are the most
entropic entities of the Universe \cite{pavon}. This brief
consideration spurs the researchers to explore the thermodynamic
consequences of phantom - dominated Universes. In doing so one must
take into account that ever accelerating Universes have a future
event horizon (or cosmological horizon) \cite{pavon}. The
thermodynamical properties associated with the apparent horizon have
been found in a quasi-de Sitter geometry of inflationary Universe
\cite{abdalla}. Setare \cite{setare1} considered the interacting
holographic model of dark energy to investigate the validity of the
generalized second laws of thermodynamics in non-flat (closed)
Universe enclosed by the event horizon. In \cite{sari}, it is shown
that GSL is generally valid for a system of dark energy interacting
with dark matter and radiation in FRW Universe. Further in Horava
Lifshitz cosmology, it has been shown that under detailed balance
the generalized second law is generally valid for flat and closed
geometry and it is conditionally valid for an open Universe, while
beyond detailed balance it is only conditionally valid for all
curvatures \cite{sari1}. In a comprehensive review, the GSL has been
extended in various generalized gravity theories including Lovelock,
Gauss-Bonnet, braneworld, scalar-tensor and $f(R)$ models
\cite{wang}.

In Einstein's gravity, the entropy of the horizon is proportional to
the area of the horizon, $S\propto A$. When gravity theory is
modified by adding extra curvature terms in the action principle, it
modifies to the entropy-area relation, for instance, in $f(R)$
gravity, the relation is $S\propto f'(R)A$ \cite{Wald}. On the other
hand, quantum corrections to the semi-classical entropy law have
been introduced in recent years, namely logarithmic and power law
corrections . Logarithmic corrections, arises from loop quantum
gravity due to thermal equilibrium fluctuations and quantum
fluctuations \cite{Meis, Ghosh, Chatt, Ban, Mod, Jamil1, Sad}
\begin{equation}
S= \frac{A}{4}+\alpha\ln\frac{A}{4}.
\end{equation}
On its part, power law corrections appear in dealing with the
entanglement of quantum fields in and out the horizon \cite{Rad,
She, Das}
\begin{equation}
S= \frac{A}{4}\left[1-K_{\alpha}A^{1-\frac{\alpha}{2}}\right],
\end{equation}
where,
\begin{equation}
A=4\pi
R_{h}^{2}~;~~~~K_{\alpha}=\frac{\alpha(4\pi)^{\frac{\alpha}{2}-1}}{(4-\alpha)r_{c}^{2-\alpha}}.
\end{equation}
Here, $r_{c}$ is the crossover scale and $\alpha$ is the
dimensionless constant whose value is currently under debate. The
second term in Eq. (2) can be considered as a power-law correction
to the entropy-area law, arising from entanglement of the
wave-function of the scalar field between the ground state and the
exited state \cite{Rad,She,Das}. The correction term is also more
significant for higher excitations. It is important to note that the
correction term decreases faster with $A$ and hence in the
semi-classical limit (large area) the entropy-area law is recovered.

The plan of the paper is as follows: In section II, we write down
basic equations of cosmology for our further use. In section III, we
investigate the GSL with power-law entropy correction for both
apparent horizon (subsection - A) and future event horizon
(subsection - B). In section IV, we will discuss the GSL at both
horizons in the non-equilibrium setting. Finally we conclude this
paper.

\section{\bf Basic equations}

We consider a homogeneous and isotropic spatially flat ($k=0$) FRW
Universe which is described by the line element
\begin{equation}
ds^{2}=-dt^{2}+a^{2}(t)\left[dr^{2}+r^{2}(d\theta^{2}+\sin^{2}\theta
d\phi^{2}) \right].
\end{equation}
Now assume that the Universe is filled with a perfect fluid of
n-components (such as dark energy, dark matter, radiation and so
on): $\rho=\sum_{i=1}^{n}\rho_{i}$ and $p=\sum_{i=1}^{n}p_{i}$ where
$\rho$ and $p$ are total energy density and pressure of the combined
fluid. So the Einstein's field equations are given by
\begin{equation}
H^{2}=\frac{8\pi}{3}~\rho,
\end{equation}
and
\begin{equation}
\dot{H}=-4\pi(\rho+p).
\end{equation}
The energy conservation equation is given by
\begin{equation}
\dot{\rho}+3H(\rho+p)=0.
\end{equation}
Now consider there is an interaction between all fluid components
\cite{zim}. We cannot specify the form of interaction between the
components as the nature of dark energy and dark matter are not
understood yet. These interactions modify the form of the equation
of states of the interacting fluids. Interest in these models has
been spurred when it was found that the evolution of the universe,
from early deceleration to late time acceleration can be explained
\cite{bin}. In addition such an interacting dark energy model can
accommodate a transition of the dark energy from a quintessence
state $w_D>-1$ to $w_D<-1$ phantom state and explain the coincidence
problem too \cite{karami}. Observational constraints on interacting
dark energy models have been obtained and the model fits with 95\%
confidence limits with the data \cite{roy}.

The continuity equations for individual fluids are
\begin{equation}
\dot{\rho}_{i}+3H(\rho_{i}+p_{i})=Q_{i},
\end{equation}
where $Q_{i}$ is an interaction term which can be an arbitrary
function of cosmological parameters like the Hubble parameter and
energy densities \cite{sadjadi}. This term allows the energy
exchange between the components of the perfect fluid and may
alleviate the coincidence problem. From above two equations (7) and
(8), we find  $\sum_{i=1}^{n}Q_{i}=0$.\\

\section{\bf GSL of Thermodynamics with Power Law Entropy Correction}

Using Gibb's law for each component of the fluid, we have
\begin{equation}
T_{i}dS_{i}=d(\rho_{i}V)+p_{i}dV,
\end{equation}
where $T_{i}$ is the temperature and $S_{i}$ is the entropy of the
$i$-th component of the fluid. If $R_{h}$ be the radius of the
horizon then the volume can be written as $V=\frac{4}{3}~\pi
R_{h}^{3}$ (assuming spherical symmetry). From the above equation,
after simplification, we obtain
\begin{equation}
\dot{S}_{i}=\frac{4}{3}~\pi R_{h}^{3}\frac{Q_{i}}{T_{i}}+4\pi
R_{h}^{2}(\dot{R}_{h}-HR_{h})\frac{\rho_{i}+p_{i}}{T_{i}}.
\end{equation}
Now the total changes of entropy inside the horizon is given by
\begin{equation}
\dot{S}_{I}=\sum_{i=1}^{n}\dot{S}_{i}=\frac{4}{3}~\pi
R_{h}^{3}\sum_{i=1}^{n}\frac{Q_{i}}{T_{i}}+4\pi
R_{h}^{2}(\dot{R}_{h}-HR_{h})\sum_{i=1}^{n}\frac{\rho_{i}+p_{i}}{T_{i}}.
\end{equation}
In cosmological models of accelerated Universe, there are horizons
to which we can assign an entropy as a measure of information behind
them. The most natural horizon of the Universe is the apparent
horizon\footnote{It is the radius obtained by solving the equation
$g^{\mu\nu}\partial_\mu\tilde r\partial_\nu\tilde r=0$, where
$\tilde r=a(t)r$ i.e. the LHS vanishes at the apparent horizon
$R_A$. For FRW Universe, it gives
$R_A=\frac{1}{\sqrt{H^2+\frac{k}{a^2}}}$, thus $R_A=\frac{1}{H}$ for
$k=0$. } whose radius is $R_{A}=\frac{1}{H}$ (also called the Hubble
horizon). Another cosmological horizon which conceptually more
resembles to the black-hole horizon is the future event horizon,
whose radius, $R_{E}$ is defined by \cite{ibrar}
\begin{equation}
R_{E}=a(t)\int_{t}^{\infty}\frac{dt'}{a(t')}\ <\infty.
\end{equation}
It describes the distance that light travels from the present time
to an arbitrary time in future. Despite the presence of an infinity,
the horizon $R_{E}$ can be finite. For future event horizon, the
derivative of the radius of event horizon can be written as
\begin{equation}
\dot{R}_{E}=HR_{E}-1.
\end{equation}
Only in a de Sitter spacetime we have $R_{A}=R_{E}$. For generality,
in our study we consider both the choices: $R_{A}$ and $R_{E}$. Note
that cosmological event horizon does not always exist for all FRW
universes, the apparent horizon and the Hubble horizon always do
exist.

\subsection{On the apparent horizon $R_{A}$}

If we take the natural horizon of the Universe as the apparent
horizon i.e., $R_{h}=R_{A}=H^{-1}$, from the equation (2), the
entropy on the apparent horizon can be written as
\begin{equation}
S_{A}=\frac{\pi}{H^{2}}\left[1-\frac{\alpha}{4-\alpha}(r_{c}H)^{\alpha-2}\right].
\end{equation}
The rate of change of the entropy on the apparent horizon (using
(14)) is obtained as
\begin{equation}
\dot{S}_{A}=-\frac{2\pi
\dot{H}}{H^{3}}\left[1-\frac{\alpha}{2}\left(r_{c}H\right)^{\alpha-2}\right].
\end{equation}
From (11), we obtain the rate of change of the entropy inside the
apparent horizon as
\begin{equation}
\dot{S}_{I}=\frac{4\pi}{3H^{3}}\sum_{i=1}^{n}\frac{Q_{i}}{T_{i}}-
4\pi\left(\frac{\dot{H}}{H^{4}}+\frac{1}{H^{2}}\right)\sum_{i=1}^{n}\frac{p_{i}+\rho_{i}}{T_{i}}.
\end{equation}
Hence adding (15) and (16), we have the rate of change of total
entropy as
\begin{equation}
\dot{S}=\dot{S}_{I}+\dot{S}_{A}=\frac{4\pi}{3H^{3}}\sum_{i=1}^{n}\frac{Q_{i}}{T_{i}}
-4\pi\left(\frac{\dot{H}}{H^{4}}+\frac{1}{H^{2}}\right)\sum_{i=1}^{n}\frac{p_{i}+\rho_{i}}{T_{i}}-\frac{2\pi
\dot{H}}{H^{3}}\left[1-\frac{\alpha}{2}\left(r_{c}H\right)^{\alpha-2}\right].
\end{equation}

$\bullet{}$ {\bf GSL in Thermal Equilibrium:}

In {\it thermal equilibrium}, we have $\forall i:T_{i}=T$ i.e. the
horizon temperature matches with the temperature of n-component
fluid. In the present state of the Universe, this general assumption
is not so valid since radiation temperature is higher compared to
non-relativistic cold dark matter. However thermal equilibrium did
occur in the early radiation dominated Universe when all
energy-matter was in thermal equilibrium with the radiation.  When
the horizon is the apparent horizon, we take the temperature as the
Hawking temperature $T=\frac{H}{2\pi}$, after simplification, we get
from (17) and (6) that
\begin{equation}
\dot{S}=\frac{2\pi
\dot{H}}{H^{3}}\left[\frac{\alpha}{2}\left(r_{c}H\right)^{\alpha-2}+\frac{
\dot{H}}{H^{2}}\right].
\end{equation}
It may be derived that GSL would hold if
\begin{equation}
\dot{S}\ge 0, \ \ \ \Rightarrow \ \ \
\dot{H}\left[\frac{\alpha}{2}\left(r_{c}H\right)^{\alpha-2}+\frac{
\dot{H}}{H^{2}}\right] \ge 0.
\end{equation}

From above we see that the GSL is always true for $\alpha=0$ (for
any sign of $\dot{H}$). We investigate two interesting case here:

\begin{itemize}

\item For quintessence dominated era, $\dot{H}<0$, the GSL holds (a)
for all $\alpha<0$ and (b) for $\alpha>0$ with
$\dot{H}<-\frac{\alpha}{2}\left(r_{c}\right)^{\alpha-2}H^{\alpha}$
but GSL breaks down for $\alpha>0$ with
$0>\dot{H}>-\frac{\alpha}{2}\left(r_{c}\right)^{\alpha-2}H^{\alpha}$.
\item For phantom dominated era, $\dot{H}>0$, the GSL holds (a) for
all $\alpha>0$ and (b) for $\alpha<0$ with
$\dot{H}>-\frac{\alpha}{2}\left(r_{c}\right)^{\alpha-2}H^{\alpha}$
but GSL breaks down for $\alpha<0$ with
$0<\dot{H}<-\frac{\alpha}{2}\left(r_{c}\right)^{\alpha-2}H^{\alpha}$.
Equation (19) puts some constraint on $\alpha$.

\end{itemize}

At this juncture it would be investigated whether GSL remains
valid in the case of small perturbations around the de Sitter
space (quasi-de-Sitter spacetime). As an illustration we consider,

$$H=H_{0}+H_{0}^{2}\epsilon t+\mathcal{O}(\epsilon^{2});
\epsilon=\frac{\dot{H}}{H^{2}}; \mid\epsilon\mid\ll1.$$

(i) When $\dot{H}>0$ which corresponds to super-acceleration (or
phantom) phase, we may conclude that the GSL is satisfied when
\begin{equation}
\alpha\geq-2\epsilon \left(r_{c}H\right)^{2-\alpha}\Rightarrow
\alpha\left(r_{c}H\right)^{2}\geq-2\epsilon
\left(r_{c}H\right)^{\alpha}.
\end{equation}

(ii) When $\dot{H}<0$ (quintessence phase), we may conclude that the
GSL is satisfied if
\begin{equation}
\alpha\leq-2\epsilon \left(r_{c}H\right)^{2-\alpha}\Rightarrow
\alpha\left(r_{c}H\right)^{2}\leq-2\epsilon
\left(r_{c}H\right)^{\alpha}.
\end{equation}

Two constraints on $\alpha$ are now available in order GSL to valid
via (20) and (21) in the  case of small perturbations around the de
Sitter space. Note that precise value of $\alpha$ can not be
determined via the inequalities. However observational constraints
could be helpful for this purpose but that is out of scope of this
paper.

\subsection{On the event horizon $R_{E}$}

In this section, the validity of GSL would be investigated on the
event horizon. From equation (2), we obtain the entropy on the
event horizon as
\begin{equation}
S_{E}=\pi
R_{E}^{2}\left[1-\frac{\alpha}{4-\alpha}\left(\frac{R_{E}}{r_{c}}\right)^{2-\alpha}\right].
\end{equation}
Differentiating (22) w.r.t. time $t$, we obtain
\begin{equation}
\dot{S}_{E}=2\pi
R_{E}(HR_{E}-1)\left[1-\frac{\alpha}{2}\left(\frac{R_{E}}{r_{c}}\right)^{2-\alpha}\right].
\end{equation}
Adding (11) and (23), the rate of change of total entropy for event
horizon is obtained as
\begin{equation}
\dot{S}=\dot{S}_{I}+\dot{S}_{E}=\frac{4\pi}{3H^{3}}\sum_{i=1}^{n}\frac{Q_{i}}{T_{i}}-4\pi
R_{E}^{2}\sum_{i=1}^{n}\frac{p_{i}+\rho_{i}}{T_{i}}+2\pi
R_{E}(HR_{E}-1)\left[1-\frac{\alpha}{2}\left(\frac{R_{E}}{r_{c}}\right)^{2-\alpha}\right].
\end{equation}
The GSL requires $\dot{S}\geq 0$, i.e. the sum of the entropies of
the perfect fluids inside the event horizon and the entropy
attributed to the horizon is a non-decreasing function of the
comoving time. In the following we discuss the validity of this
law specially in the presence of dark energy.\\

$\bullet{}$ {\bf GSL in Thermal Equilibrium:} In thermal
equilibrium, we have $\forall i:T_{i}=T$. So equation (24) becomes
\begin{equation}
\dot{S}=\frac{\dot{H}}{T}R_{E}^{2}+2\pi
R_{E}(HR_{E}-1)\left[1-\frac{\alpha}{2}\left(\frac{R_{E}}{r_{c}}\right)^{2-\alpha}\right].
\end{equation}
For the future event horizon, and in the absence of a well-defined
temperature, we assume that $T$ is proportional to the Hawking
temperature \cite{davies}
\begin{equation}
T=\frac{bH}{2\pi},
\end{equation}
where $b$ is an arbitrary constant of order unity. Equation (25)
yields
\begin{equation}
\dot{S}=\frac{2\pi \dot{H}}{bH}R_{E}^{2}+2\pi
R_{E}(HR_{E}-1)\left[1-\frac{\alpha}{2}\left(\frac{R_{E}}{r_{c}}\right)^{2-\alpha}\right].
\end{equation}
GSL will be satisfied if
\begin{equation}
\dot{S}\ge 0 \ \  \Rightarrow \ \
\frac{2r_{c}^{2}}{R_{E}(HR_{E}-1)}~\frac{d}{dt}\log(R_{E}H^{\frac{1}{b}})
\ge \alpha\left(\frac{r_{c}}{R_{E}}  \right)^{\alpha}.
\end{equation}

As the expression (28) appears to be very complicated and it is not
possible to draw any definite conclusion regarding the parameters of
the model, we consider a particular choice of scale factor that
pertains to a phantom dominated Universe of polelike type described
by \cite{nojiri}
\begin{equation}
a(t)=a_{0}(t_{s}-t)^{-n}~~~~~~a_{0}>0;~~n>0;~~t_{s}>t.
\end{equation}
 For this choice of scale factor, we have
\begin{equation}
H=\frac{n}{t_{s}-t}~;~~R_{E}=\frac{t_{s}-t}{n+1}~;~~\dot{H}=\frac{n}{(t_{s}-t)^{2}}.
\end{equation}
So equation (27) reduces to the form
\begin{equation}
\dot{S}=\frac{2\pi
(t_{s}-t)}{(n+1)^{2}}\left[-1+\frac{1}{b}+\frac{\alpha}{2}\left(\frac{t_{s}-t}
{r_{c}(n+1)}\right)^{2-\alpha}\right].
\end{equation}
The GSL is satisfied if $\dot{S}\ge 0$ i.e.,
\begin{equation}
t_{s}-t \ge  r_{c}(n+1)
\left[\frac{2(b-1)}{b\alpha}\right]^{\frac{1}{2-\alpha}}~~~\text{with}~~b>1.
\end{equation}

Combining the expressions (29) and (32), we get an upper limit of
the scale factor $a(t)$ as

$$ a(t)\le \frac{a_{0}}{\{r_{c}(n+1)\}^{n}}\left[\frac{2(b-1)}{b\alpha}\right]
^{\frac{n}{\alpha-2}} .$$

Therefore the GSL would be valid if the scale factor lies below
the r.h.s of the above expression.\\

\section{ GSL in Thermal Non-Equilibrium}

If we drop the condition of thermal-equilibrium, the problem of
investigating the validity of the GSL becomes more complicated. This
situation arises since dark matter, radiation and dark energy have
different temperatures \cite{sol}. In this case, the components of
temperatures $T_{i}$'s are all distinct. The equation (6) can be
written as
\begin{equation}
\dot{H}=-4\pi\sum_{i=1}^{n}(\rho_{i}+p_{i}).
\end{equation}
Now
\begin{equation}
\sum_{i=1}^{n}\left(\frac{\rho_{i}+p_{i}}{T_{i}}\right)=\frac{p_{1}+\rho_{1}}{T_{1}}
+\sum_{i=2}^{n}\left(\frac{\rho_{i}+p_{i}}{T_{i}}\right)=-\frac{\dot{H}}{4\pi
T_{1}} -\frac{1}{T_{1}}\sum_{i=2}^{n}(\rho_{i}+p_{i})+\sum_{i=2}^{n}
\left(\frac{\rho_{i}+p_{i}}{T_{i}}\right).
\end{equation}
We assume a particular choice for $T_{1}=\frac{bH}{2\pi}$ and the
other $T_{i}$'s are considered arbitrarily. Furthermore,
considering $p_{i}=w_{i}\rho_{i}$ for $i=2,3,....,n$ we obtain
\begin{equation}
\sum_{i=1}^{n}\left(\frac{\rho_{i}+p_{i}}{T_{i}}\right)=-\frac{\dot{H}}{2bH}
-\sum_{i=2}^{n}\left[(1+w_{i})\rho_{i}\left\{\frac{2\pi}{bH}-
\sum_{i=2}^{n}\frac{1}{T_{i}}\right\}\right].
\end{equation}
In this model $\rho_{1}$ and $\{\rho_{2},\rho_{3},...,\rho_{n}\}$
consist of ingredients in thermal equilibrium (in each subset),
which even if do not interact with the elements of the other
subset, can interact with each other. Using (35) in (17) we get
that for GSL to be valid on the apparent horizon $R_{A}$ we
require
\begin{equation}
\dot{S}=\frac{\pi}{br^{2}_{c}H^{5}}\left[2\dot{H}r^{2}_{c}(\dot{H}-(-1+b)H^{2})
+b\alpha(r_{c}H)^{\alpha}\dot{H}+4bH(\dot{H}+H^{2})r^{2}_{c}\sum_{i=2}^{n}(1+w_{i})\rho_{i}
\left\{\frac{2\pi}{bH}-\sum_{i=2}^{n}\frac{1}{T_{i}}\right\}\right]\geq0.
\end{equation}
Using (35) in (24) we get that for GSL to be valid on the event
horizon $R_{E}$ we require
\begin{equation}
\dot{S}=2\pi
R_{E}\left[(HR_{E}-1)\left(1-\frac{\alpha}{2}\left(\frac{R_{E}}{r_{c}}\right)^{2-\alpha}\right)
+\frac{R_{E}}{bH}\left\{\dot{H}+2bH\sum_{i=2}^{n}(1+w_{i})\rho_{i}\left(\frac{2\pi}{bH}
-\sum_{i=2}^{n}\frac{1}{T_{i}}\right)\right\}\right]\geq0.
\end{equation}

From the above expressions (36) and (37) we get general conditions
for the validity of GSL for apparent and event horizons in case of
thermal non-equilibrium. However, it is not possible to obtain any
specific constraints on the model parameters for the validity of
GSL.

\section{\bf Conclusion}

In this paper, we considered the FRW cosmological spacetime composed
of interacting components. We discussed the generalized second law
of thermodynamics by considering the power law correction of entropy
on the horizon. We focused on both the apparent and future event
horizons and expressed the time derivatives of the total entropies
in terms of the model parameters $\alpha$ and $r_{c}$. Considering
the GSL in thermal equilibrium on the apparent horizon we find from
equation (19) the conditions for validity of the GSL on the
quintessence and phantom dominated era. Also, considering a small
perturbation around the de Sitter space-time we found the conditions
on the model parameters required for the validity of GSL.
Considering the GSL in thermal equilibrium on the event horizon we
find that the GSL is valid if
$\left(\frac{t_{s}-t}{r_{c}(n+1)}\right)^{2-\alpha} \ge
\frac{2(b-1)}{b\alpha}$ (equation (32)). In equation (37) we
expressed the time derivative of total entropy in terms of $\alpha$
and $r_{c}$. If we consider a small perturbation around the de
Sitter space-time, the general conditions of the validity of GSL
have been found. Also if a phantom dominated Universe has a polelike
type scale factor, the validity of GSL has also been analyzed.

\subsection*{Acknowledgment}
The authors would like to thank anonymous referee for giving useful
comments to improve this work.

\end{document}